\documentclass[prl,english,graphics,showpacs,twocolumn,superscriptaddress]{revtex4-1}
\usepackage{graphicx}
\usepackage{subfig}
\usepackage{amssymb,amsmath,amsfonts,mathrsfs,eufrak,fancyhdr}
\usepackage{hyperref}
\usepackage{color}
\usepackage{makeidx,endnotes}
\usepackage{showidx}
\usepackage{theorem,eepic}
\usepackage{epsf,epsfig}

\def\vec#1{{\boldsymbol{#1}}}
\def\<{\langle}
\def\>{\rangle}

\def\set#1{{\sf #1}}

\def\map#1{{\mathcal{#1}}}

\def\Tr{\operatorname{Tr}}

\def\diag{\operatorname{diag}}

\begin{document}
\title{Entanglement and decoherence: fragile and robust entanglement}
\author{Jaroslav Novotn\'y}
\affiliation{Department of Physics, FNSPE, Czech Technical University in Prague, B\v rehov\'a 7, 115 19 Praha 1 - Star\'e M\v{e}sto, Czech Republic}
\affiliation{Institut f\"ur Angewandte Physik, Technische Universit\"at Darmstadt, D-64289 Darmstadt, Germany}
%\email{jaroslav.novotny@seznam.cz}
\author{Gernot Alber}
\affiliation{Institut f\"ur Angewandte Physik, Technische Universit\"at Darmstadt, D-64289 Darmstadt, Germany}
\author{Igor Jex}
\affiliation{Department of Physics, FNSPE, Czech Technical University in Prague, B\v rehov\'a 7, 115 19 Praha 1 - Star\'e M\v{e}sto, Czech Republic}

\begin{abstract}
The destruction of entanglement of open quantum systems by decoherence is investigated in the asymptotic long-time limit.
Starting from a general and analytically solvable decoherence model which does not involve any weak-coupling or Markovian assumption
it is shown that two fundamentally different classes of entangled states can be distinguished. Quantum states of the first class are fragile against decoherence so that they can be disentangled asymptotically even if coherences between pointer states are still present. Quantum states of the second type are robust against decoherence. Asymptotically they can be disentangled only if also decoherence is perfect. A simple criterion for identifying these two classes on the basis of two-qubit entanglement is presented.
\end{abstract}
\maketitle

%{\it Introduction.---}
Every natural physical quantum object is in contact with an environment.
A consequence of the resulting inevitable interaction of such an open quantum system with its environment
is decoherence \cite{Joos_book,Schlosshauer,Zurek2003}.
For sufficiently long interaction times
this process leads to an environment-induced selection of so-called pointer states
which remain at least approximately undisturbed while any of their superposition states decays quickly into a mixture.
This asymptotic loss of quantum coherence is responsible for the appearance of classical features of open quantum systems
\cite{Joos_book,Schlosshauer,Zurek2003}.
Apart from that decoherence is also a process which tends to destroy entanglement.
Nevertheless, the recently discussed time-dependent phenomenon of sudden death of entanglement \cite{Yu2003}
has already hinted at some subtle relations between decoherence and disentanglement
in the time domain.
Thereby, it has been observed that entanglement of an open quantum system may already disappear
after a finite interaction time with its environment  while complete decoherence requires an infinite amount of time.
A further exploration of the subtleties between entanglement and decoherence
is not only of fundamental physical interest but is also of significance
for practical applications in the area of
quantum information science \cite{Nielsen} in which entanglement is a key resource.
For the realization of large quantum information processors, for example, it is of vital importance to
understand how decoherence-induced loss of entanglement of an open quantum system scales with the size of an environment
in the limit of arbitrarily long interaction times.
Under which conditions are
already finitely sized environments capable of disentangling open quantum systems completely?
Are there classes of entangled quantum states whose properties differ significantly under decoherence?
It is a main purpose of this letter to explore these questions.

%{\it A decoherence model.---}
For this purpose
a sufficiently general model of decoherence is needed
which is exactly solvable without further simplifying assumptions concerning, e.g., sizes of environments,
initially prepared quantum states, interaction strengths, or correlation times
between system and environment. Of particular interest are
generic asymptotic long-time properties of such a general decoherence model which exhibit clearly
the intricate interplay between decoherence and destruction of entanglement.

\begin{figure}[h]
\includegraphics[width=2cm]{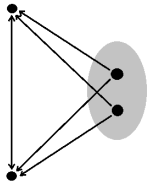}
\caption{Schematic representation of a system-environment qubit network:
$k=2$ qubits of the open quantum system (grey region) interact with $n=2$ environmental qubits.
The directed edges  indicate all possible elementary two-qubit couplings.
The tail (head) of an edge symbolizes the control (target) qubit.}
\label{fig1}
\end{figure}

Motivated by the practical significance of elementary distinguishable two-level systems (qubits) for purposes of
quantum information processing in the following we present a qubit-based class of such decoherence models.
Thereby, we start from the recent observation  that decoherence of a single system qubit
can be modeled by a sequence of 'controlled-U' unitary transformations
in which the single system qubit acts as a control and the environmental qubits act as targets \cite{Ziman2005}.
This result was obtained under the simplifying assumptions that the environment is formed by an infinite number
of qubits, that each qubit is prepared initially in the same state so that all environmental qubits are uncorrelated,
that the environmental qubits do not interact among themselves, and that each environmental target qubit interacts
with the single system qubit only once.
In order to overcome all these restrictive approximations let us consider as a natural generalization
$k$ system qubits which interact with  $n$ environmental qubits
by a sequence of elementary interactions or 'collisions'. All of these 'collisions'
are assumed to be well separated in time.
In each of them a pair of qubits, i.e. a control qubit $i$ and a target qubit $j$,
is selected randomly and
a controlled unitary transformation
\begin{eqnarray}
\label{definition_c_rotation}
\hat{U}^{(\phi)}_{ij} &=& |0\>_{ii}\<0|\otimes \hat{I}_j +
|1\>_{ii}\<1| \otimes \hat{u}^{(\phi)}_j
\end{eqnarray}
is applied.
Thereby,
$\hat{I}_j$ is the unit operator acting on qubit $j$ and
\begin{eqnarray}
\label{condunit}
\hat{u}^{(\phi)}_j &=&
\cos\phi (|0\>_{jj}\<0| - |1\>_{jj}\<1|) + \sin\phi (|0\>_{jj}\<1| + |1\>_{jj}\<0|)\nonumber\\
&&
\end{eqnarray}
denotes the unitary one-qubit transformation acting on the target qubit $j$.
Such controlled unitary couplings have already been investigated in the context of various decoherence models
\cite{Joos_book,Zurek2003}
because they do not affect the computational basis states $|0\>_i$ and $|1\>_i$
of the control qubit $i$ and at the same time they can decrease any quantum coherence between these two system states.
Within our generalized decoherence model
each system qubit $i$ is a possible control qubit for an elementary interaction $\hat{U}^{(\phi)}_{ij} $
with any environmental target qubit $j$.  Furthermore, all environmental qubits also interact among themselves.
However, in order to guarantee a well defined pointer basis in our decoherence
model we assume that the system qubits do not interact among themselves.
The interaction pattern characterizing which qubits can be coupled by an elementary interaction
$\hat{U}^{(\phi)}_{ij} $ between randomly chosen qubits $i$ and $j$
can be encoded in a convenient way in a weighted and directed interaction graph
\cite{Novotny2011}.
Its set of vertices represents the $n+k$ qubits of the total qubit system and the set of directed edges $\set E$
represents the possible interactions among the qubits.
Each of these edges $e=ij \in \set E$ is weighted with a probability $p_e$ with which
qubits $i$ and $j$ are coupled by
the unitary two-qubit transformation $\hat{U}^{(\phi)}_{e}$.
These probabilities are normalized to unity, i.e. $\sum_{e\in \set E} p_e = 1$.
A simple example of such a qubit network
is depicted in Fig. \ref{fig1}.

Thus, within this decoherence model the quantum state $\hat{\rho}(N)$ resulting from $N$
completed elementary interactions
is changed by the $(N+1)$-th elementary interaction
to the quantum state
\begin{equation}
\label{one_step}
\hat{\rho}(N+1)=\sum_{e \in \set E} p_e \hat{U}^{(\phi)}_e
\hat{\rho}(N)
\hat{U}^{(\phi)\dagger}_e \equiv \map P(\hat{\rho}(N)).
\end{equation}
The map $\map P$
is a random unitary operation \cite{Bengtsson2006} and
describes an elementary interaction or 'collision' between system and environment.
General properties of quantum states $\hat{\rho}(N)$
resulting from repeated applications of such quantum maps
have been studied recently \cite{Novotny2010}.
In particular,
it has been shown that in the asymptotic limit of $N\gg1$ the
quantum state
$\hat{\rho}(N)$
becomes independent of the probability distribution $\{p_e, e\in E\}$ and that it
is determined uniquely by a linear attractor space. This latter space is formed by the maximal set
of all possible orthonormal solutions $\hat{X}_{\lambda,i}$ of the eigenvalue equations
\begin{eqnarray}
\label{commutation_relations}
\hat{U}^{(\phi)}_{e} \hat{X}_{\lambda,i}\hat{U}^{{(\phi)}\dagger}_e
&=&\lambda \hat{X}_{\lambda,i}~~{\rm for~all}~e\in \set E
\end{eqnarray}
with $|\lambda|=1$.
The index $i$ distinguishes different mutually orthonormal solutions with eigenvalue $\lambda$.
If these attractors are known, in the asymptotic limit $N \gg 1$ the quantum state
resulting from an initially prepared state $\hat{\rho}_{in}$
is given by
\begin{eqnarray}
\label{longtime dynamics}
\hat{\rho}(N) &=& \map P^{N}(\hat{\rho}_{in})
 = \sum_{\mid\lambda \mid =1,i}
\lambda^N\Tr\{\hat{\rho}_{in}\hat{X}_{\lambda,i}^{\dagger}\} \hat{X}_{\lambda,i}.
\end{eqnarray}

With the help of the methods developed in Ref. \cite{Novotny2011}
it can be shown
that within our decoherence model for $n\geq 2$ the only possible eigenvalue
of Eqs.(\ref{commutation_relations})
is given by
$\lambda=1$ and
its eigenspace is spanned by the
$4^k+3\times2^k+1$ linear independent solutions
\begin{eqnarray}
&&|\vec x\>\<\vec x| \otimes \hat{I}_n, |\vec 0\>\<\vec x| \otimes |\vec 0_n\>\<\phi_n|, |\vec x\>\<\vec 0_k| \otimes  |\phi_n\>\<\vec 0 _n|, \nonumber \\
&&|\vec x\>\<\vec y| \otimes |\phi_n\>\<\phi_n|, |\vec 0_k\>\<\vec 0_k| \otimes |\vec 0_n\>\<\vec 0_n|
\label{attractor}
\end{eqnarray}
with the pure quantum states
$|\vec{0}_n\rangle =|0\>^{\otimes n}$, $|\phi\>= \cos(\phi/2)|0\>+\sin(\phi/2)|1\>$, $|\phi_n\>= |\phi\>^{\otimes n}$.
Vectors $|\vec x\>$ and $|\vec y\>$, with $\vec x, \vec y \in \set 2^k$, are arbitrary elements of the computational basis of $k$ qubits.
Note that the quantum state $|\phi\>$ is not affected by the unitary transformation  $\hat{u}^{\phi}$ of Eq.(\ref{condunit}),
i.e. $\hat{u}^{\phi}|\phi\> = |\phi\>$.

In order to investigate the decoherence process affecting the $k$ system qubits in the
asymptotic limit of large numbers of elementary interactions $N\gg 1$ let us
consider an initially prepared quantum state
$\hat{\rho}_{in}=\hat{\rho}^{(S)} \otimes \hat{\rho}^{(E)}$
which does not contain any correlations between system $S$ and environment $E$.
Using the attractor space of Eq.(\ref{attractor})
and tracing out the environment in Eq.(\ref{longtime dynamics})
the asymptotic quantum state of the open quantum system $S$ is given by
\begin{equation}
\label{asymptotic_state}
\hat{\rho}^{(S)}_{\infty} = \diag(\hat{\rho}^{(S)}) +
\sum_{\vec x,\vec y \in \set 2^k, \vec x \neq \vec y}
\<\phi_n |\hat{\rho}^{(E)}|\phi_n\>\hat{\rho}^{(S)}_{\vec x,\vec y}|\vec x\>\<\vec y|
\end{equation}
with $\diag(\hat{\rho}^{(S)})$ denoting the diagonal part of the system state $\hat{\rho}^{(S)}$ with respect to the
pointer basis $\{|\vec x\>, \vec x \in \set 2^k\}$.
At this point it is worth emphasizing
that in view of the general results of
Ref.\cite{Novotny2011}
some of the conditions leading to Eq.(\ref{asymptotic_state})
can be relaxed. Thus, it turns out, for example, that Eq.(\ref{asymptotic_state}) is already valid if at least one pair of
environmental qubits is interacting.

According to Eq.(\ref{asymptotic_state}) the influence of the environment $E$ on the asymptotic quantum state of the system $S$
is described by the decoherence factor
$0\leq r= \<\phi_n |\hat{\rho}^{(E)}|\phi_n \> \leq 1$
which depends only on environmental properties, such as its size $n$ and the initially prepared quantum state
$\hat{\rho}^{(E)}$.
In particular, three different cases can be distinguished.
If all environmental qubits are initially prepared in the eigenstate $|\phi\>$ of the interaction $\hat{u}^{\phi}$ of
Eq.(\ref{condunit}), i.e. $r=1$, the system state $\hat{\rho}^{(S)}$ remains unchanged.
If the environmental
state $\hat{\rho}^{(E)}$ has zero overlap with quantum state $|\phi_n\>$, i.e. $r=0$,
we observe perfect decoherence in the asymptotic limit $N\gg 1$.
In all other cases asymptotic decoherence results in an only partial suppression of coherences
between the pointer states of the open system $S$.
In particular,
if the initial state of the environment is a factorized state $\hat{\xi}^{\otimes n}$, for example,
the decoherence factor is given by
$r=\<\phi_1|\hat{\xi}|\phi_1\>^{n}$. Thus, coherences between pointer state of the system $S$
decrease exponentially with increasing number $n$ of environmental qubits.
In this case perfect decoherence can only be obtained for an infinite environment.
Furthermore,
if environmental qubits share initial correlations even in the case of an environment with infinitely many qubits
perfect asymptotic decoherence cannot always be achieved.
Let us consider an initially prepared pure environmental state of the form
$\hat{\rho}_{env}=|\chi_n\>\<\chi_n|$ with $|\chi_n\>=\cos\alpha |\phi_n\> + \sin\alpha |\nu_n\>$ and $\<\phi_n|\nu_n\> = 0$,
for example. The associated decoherence factor is given by
$r= \cos^2(\alpha)$ and is thus independent of the number of environmental qubits $n$.

Let us now investigate how this decoherence process destroys entanglement in the asymptotic limit of many elementary interactions
$N\gg 1$ between system and environment.
In order to avoid problems with appropriate measures of entanglement let us
focus on an arbitrary two-qubit subsystem of an open quantum system with $k\geq 2$ system qubits and
$n$ environmental qubits.
Thus,
the entanglement of this two-qubit subsystem can be quantified by its concurrence \cite{Wootters}.

Consider first of all an initially prepared pure state $\hat{\rho}_1^{(S)} = |\psi_1\>\<\psi_1|$
of these two system qubits
with
$|\psi_1\>=a|00\>+ b|11\>$ and $|a|^2+|b|^2 =1$.
For an arbitrary environmental state $\hat{\rho}^{(E)}$ the
concurrence $C$ of the resulting asymptotic quantum state (\ref{asymptotic_state}) is given by
\begin{equation}
\label{eq_concurrence_2}
C(\hat{\rho}^{(S)}_{1\infty}) = 2|ab|\<\phi_n|\hat{\rho}^{(E)}|\phi_n\>=C(\hat{\rho}^{(S)}_{1})r.
\end{equation}
Thus,
the resulting asymptotic entanglement is equal to the initial degree of entanglement multiplied by the decoherence factor $r$.
Therefore,
in the asymptotic limit  $N\gg 1$
the disentanglement of the originally prepared entangled state $\hat{\rho}_1^{(S)} = |\psi_1\>\<\psi_1|$
is governed by the same dependencies
concerning the size $n$ and the initially prepared state of the environment $\hat{\rho}^{(E)}$
as the decoherence factor $r$.

Let us now consider
an uncorrelated state of system and environment
with
the two system qubits initially prepared in the maximally entangled state
$|\psi_2\>= 1/2(-|00\>+|01\>+|10\>+|11\>)$.
For
an arbitrary environmental state $\hat{\rho}^{(E)}$ the
concurrence $C$ of the resulting asymptotic quantum state (\ref{asymptotic_state})
is given by
\begin{equation}
\label{eq_concurrence_3}
C(\hat{\rho}^{(S)}_{2\infty})=\frac{1}{2}\max\left\{0,3\<\phi_n|\hat{\rho}^{(E)}|\phi_n\>-1\right\}.
\end{equation}
Thus, the resulting asymptotic entanglement is drastically different from the previous case.
As soon as the
decoherence factor decreases below the threshold value of $1/3$
the degree of entanglement drops to zero in the asymptotic limit $N\gg 1$.
The entanglement of the open quantum system is now
very sensitive to the size $n$ and to the state $\hat{\rho}^{(E)}$ of the environment.
In particular, consider an $n$-qubit environment initially prepared in a factorized quantum
state $\hat{\rho }^{(E)}=\hat{\xi}^{\otimes n}$ with $\<\phi_n|\hat{\rho}^{(E)}|\phi_n\> < 1$.
This implies the relation
$\<\phi_n|\hat{\rho}^{(E)}|\phi_n\> = \<\phi|\hat{\xi}|\phi\>^n$.
Therefore, as soon as the number $n$ of qubits in the environment exceeds the critical value
of
\begin{equation}
n_{sep} = \lceil -\frac{\ln{3}}{\ln{\<\phi_1|\hat{\xi}|\phi_1\>}} \rceil,
\end{equation}
the asymptotic entanglement of the system state
vanishes.
($\lceil x \rceil$ denotes the largest integer less or equal to $x$.)
This is in extreme contrast to the previous case studied in Eq.(\ref{eq_concurrence_2})
in which
entanglement decreases exponentially with the size $n$ of the environment
and never vanishes completely for any finite size of the environment.

\begin{figure}[h]
\includegraphics[width=6cm]{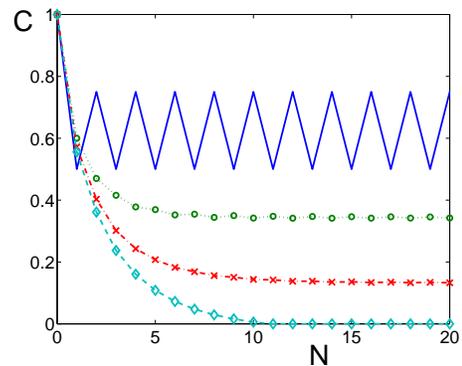}
\caption{Dependence of the concurrence $C$ of the two-qubit state $\hat{\rho}^{(S)}(N)$ on the number of elementary
interactions $N$ and on the
size of the environment $n$ ($n=1$ full, $n=2$ dotted, $n=3$ dashed-dotted, $n=4$ dashed):
The initial
state of the environment (system) is given by
$|\delta\>^{\otimes n}$ with $|\delta\>=\sin(\pi/6)|0\>+\cos(\pi/6)|1\>$ ($|\psi_2\>$)
and $\hat{u}^{(\phi)}$ is given by Eq.(\ref{condunit}) with $\phi=2\pi/3$.}
\label{fig2}
\end{figure}

The numerical simulations depicted in Fig. \ref{fig2} show
the dependence of the
concurrence $C$ of the two qubits initially prepared in the quantum
state $|\psi_2\>$ on the number $N$ of elementary interactions between system and environment. The asymptotic limit
is already achieved after a few elementary interactions. In the exceptional case of an environment of minimal size, i.e.
$n=1$, Eqs.(\ref{commutation_relations}) also allow a solution with
eigenvalue
$\lambda = -1$ which results in a non-stationary asymptotic limit leading to an oscillatory dependence.
In all other cases the asymptotic entanglement is stationary. In the case depicted in
Fig. \ref{fig2} the critical size of the environment is given by $n_{sep} = 4$ so that for all smaller sizes of the environment
asymptotically the two-qubit system is not disentangled completely.

Thus, with respect to asymptotic decoherence-induced disentanglement of an open two-qubit system
two classes of initially prepared
entangled states $\hat{\rho}^{(S)}$ can be distinguished.
The first class consists of quantum states
which are fragile against decoherence in the sense that
their asymptotic entanglement vanishes already for non-zero decoherence factors $r$
(compare with Eq.(\ref{eq_concurrence_3})).
In contrast, quantum states $\hat{\rho}^{(S)}$ of the second class
are robust against decoherence-induced
disentanglement in the sense that the two qubits
can be disentangled only if decoherence is perfect, i.e. $r=0$
(compare with Eq.(\ref{eq_concurrence_2})).
As a result these two classes of initially prepared two-qubit states
exhibit a very different behavior with respect characteristic features of the decoherence-inducing environment.

Within our general decoherence model
even a simple necessary and sufficient condition can be derived
whether the entanglement of an initially prepared
two-qubit system state $\hat{\rho}^{(S)}$ of an open system is fragile or robust.
In view of the general form of the asymptotic system state of Eq.(\ref{asymptotic_state})
the formal arguments presented in Ref.\cite{Huang2007}
lead to the conclusions that a two-qubit state
$\hat{\rho}^{(S)}$
is fragile with respect to environmental decoherence
if and only if its density matrix elements in the pointer  basis satisfy the relation
\begin{equation}
\<00|\hat{\rho}^{(S)}|00\>
\<10|\hat{\rho}^{(S)}|10\>
\<01|\hat{\rho}^{(S)}|01\>
\<11|\hat{\rho}^{(S)}|11\>
\neq 0.
\end{equation}

In summary,
a class of decoherence models has been presented by which
the intricate interplay between destruction of entanglement and decoherence
resulting from the asymptotic long-time interaction of an open qubit system with its environment
can be investigated analytically. It does not involve
any further simplifying assumptions, such as the ones involved
in Markovian models or in models in which each environmental qubit interacts only once with a system qubit.
Within this framework it has been shown that two characteristic classes of entangled quantum states can
be distinguished, namely fragile and robustly entangled states.
They exhibit significantly different dependencies on environmental properties.
For two-qubit subsystems of an open qubit system
a simple criterion for robustness and fragility has been presented.

The very existence of fragile and robustly entangled states
in open quantum systems is a consequence
of the characteristic structure of the asymptotic quantum states described by
Eq.(9).
Without the restriction $0\leq r \leq 1$ on the decoherence factor
a similar structure also appears in other recently discussed decoherence models
\cite{Joos_book,Schlosshauer,Zurek2003,Zurek2005}. Therefore, it can be conjectured that
the existence of fragile and robustly entangled quantum states is a general phenomenon accompanying any decoherence process.
In view of their significantly weaker sensitivity to decoherence
robustly entangled quantum states are expected to play a
particularly significant
role in the further development of quantum information processing and in its efforts to push
entanglement as a
characteristic quantum phenomenon
as far as possible into the macroscopic domain.

\begin{acknowledgments}
Financial support by the Alexander von Humboldt Foundation, by
CASED, and by MSM6840770039 and MSMT LC06002 of the Czech Republic
is acknowledged.
\end{acknowledgments}

\end{document}